\documentstyle[prl,aps]{revtex} \def\narrowtext{} \tighten \twocolumn
 
\begin{document}
\draft
 
\title{Bosonic Mode Mixing in the Superconducting State Spectral Function
of Bi$_2$Sr$_2$CaCu$_2$O$_{8+\delta}$}
\author{
        M. R. Norman,$^1$
        H. Ding,$^{1,2}$
        J. C. Campuzano,$^{1,2}$
        T. Takeuchi,$^{1,3}$
        M. Randeria,$^4$
        T. Yokoya,$^5$
        T. Takahashi,$^5$
        T. Mochiku,$^{6}$ and K. Kadowaki$^{6,7}$
       }
\address{
         (1) Materials Sciences Division, Argonne National Laboratory,
             Argonne, IL 60439 \\
         (2) Department of Physics, University of Illinois at Chicago,
             Chicago, IL 60607\\
         (3) Department of Crystalline Materials Science, Nagoya 
             University, Nagoya 464-01, Japan\\
         (4) Tata Institute of Fundamental Research, Bombay 400005, India\\
         (5) Department of Physics, Tohoku University, 980 Sendai, Japan\\
         (6) National Research Institute for Metals, Sengen, Tsukuba,
             Ibaraki 305, Japan\\
         (7) Institute of Materials Science, University of Tsukuba, 
             Ibaraki 305, Japan\\
         }

\address{%
\begin{minipage}[t]{6.0in}
\begin{abstract}
Photoemission spectra of Bi$_2$Sr$_2$CaCu$_2$O$_{8+\delta}$ 
below $T_c$ show two features near the $(\pi,0)$ point of the
zone:  a sharp peak at low energy and a higher binding energy hump.  We find
that the sharp peak persists at low energy even as one moves towards
$(0,0)$, while the broad hump shows significant dispersion which correlates
well with the normal state dispersion.  
We argue that these features are naturally explained
by the mixing of electrons with a bosonic mode which appears only
below $T_c$, and speculate that the latter may be related to the resonance
seen in recent neutron data.
\typeout{polish abstract}
\end{abstract}
\end{minipage}}

\maketitle
\narrowtext
Angle resolved photoemission data on the
quasi-two dimensional high temperature superconductors can be
interpreted in terms of the one-electron spectral function \cite{NK}.
This implies that important information about the self-energy $\Sigma$,
and how it changes from the normal to the superconducting (SC) state,
can, in principle, be obtained by analysis of the ARPES lineshape.  
This obviously has important ramifications in
elucidating a proper microscopic theory of high temperature superconductors.

Perhaps the most dramatic effect in this regard
is the temperature dependence of the lineshape in Bi2212. 
A very broad normal state spectrum near the $(\pi,0)$
point of the zone evolves quite rapidly for $T < T_c$ into a sharp,
resolution limited, quasiparticle peak \cite{NK}
followed at higher binding energies by a dip \cite{REVIEW,DING96}
then a hump, the latter corresponding to where
the spectrum recovers to its normal state value.
Similar effects are observed in tunneling\cite{TUNNEL}.

In this paper we focus on another remarkable difference between
the normal state and SC state data which has not been noticed earlier.
In Fig.~1, we show spectra for a $T_c=$87K Bi2212 sample
along $\Gamma-\bar{M}-Z$, 
i.e., $(0,0)-(\pi,0)-(2\pi,0)$, (a) in the normal state (105 K) and 
(b) in the SC state (13 K) from which we note two striking features.
First, we see that the low energy peak in the SC state
persists over a surprisingly large range in ${\bf k}$-space,
even when the normal state spectra have dispersed far from the Fermi energy. 
For example, the sharp peak is visible at about 40meV even in
curve 4  of Fig.~1b, when the corresponding normal state spectrum
is peaked 320 meV below $E_{F}$. 
Second, when the hump in the SC state disperses, it essentially
follows that of the normal state spectrum.

To highlight these two features, we plot in
Fig.~2 the position of the sharp peak and that of the hump
in the SC state versus the single (broad) spectral peak
position in the normal state.  
The experimental points in this figure are, by themselves,
highly suggestive of the quantum mechanical mixing of two levels
in the SC state electronic spectral function.
We note the similarity of
this plot to the spectrum of an electron interacting with a sharp 
phonon mode\cite{SCHRIEFFER} in conventional superconductors.
At the same time it is very important to stress that, in marked contrast
to the phonon case, the ``bosonic mode'' seen here 
appears {\it only} in the superconducting state, and,
as we shall argue below, has a many-body origin.

We will discuss below a phenomenological self-energy for the SC state
which incorporates the mixing of electron states 
with a bosonic mode. The resulting spectral functions
are not only able to capture the features of
the lineshape for a given ${\bf k}$, but, more
importantly, also able to describe
the very different dispersions of the sharp peak and the broad hump.
Note that, while near $k_F$  the sharp peak is quasiparticle-like
in nature, it is mostly bosonic-like once the
higher energy hump has begun to disperse.
It is likely that this bosonic mode has the same physical
origin as that recently observed by neutron scattering in YBCO\cite{YBCO}.

The data of Fig.~1 were obtained on high quality slightly overdoped 
Bi2212 single crystals ($T_c$ = 87 K), with measurements carried out at 
the Synchrotron Radiation Center, Wisconsin, 
using a high resolution 4m normal 
incidence monochromator\cite{DING95}. 
22eV photons polarized along $\Gamma-\bar{M}$ (the Cu-O bond direction) 
were used for both narrow energy scans (resolution FWHM=18 meV) 
and wide energy scans (FWHM=35 meV).  Similar results were seen on a variety
of samples with different doping levels, photon polarizations,
and photon energies.

We have already noted above the persistence of the
low frequency peak in the SC state from
curves 4 through 16 in Fig.~1b, about half of the Brillouin zone.  As one
moves from either extreme towards $\bar{M}$ (curve 11), (1) the broad
hump at high binding energy moves towards $E_F$ and (2) weight
is transferred from the hump to the low frequency peak, which is
is fairly fixed in energy.  The same phenomena are also seen
along $\bar{M}$ to $Y$ (Fig.~1c).

A potential complication in interpreting these data
is the influence of images of the CuO bands (ghosts) caused by the
incommensurate superlattice \cite{DING96,DING95}, 
since one predicts a Fermi crossing of one of
these images near curve 4.  The following arguments can be made
against such an effect.  First, the ghosts are not visible in the normal
state in this polarization geometry.  They do however become quite visible if 
the photon polarization is rotated by 45$^\circ$ \cite{DING96}. 
Moreover, comparison of superconducting state spectra in these two 
polarizations indicate that the midpoint
of the leading edge in the present polarization (20 meV) 
is near that of the $\bar{M}$ point, whereas in the 45 degree rotated 
polarization, the midpoint is 5 meV.  
The latter value is consistent with the expected
gap size on the ghost band for this k value.  Second, the intensity at the
peak position (35 meV) monotonically rises from $\Gamma$ with a maximum near
$\bar{M}$, indicating only one spectral feature.

We can also eliminate the two bilayer-split bands as the explanation
of the non-trivial lineshape with a peak, dip and hump.
First, we recall our earlier results \cite{DING96} on the intensities
of the peak and the hump at $\bar{M}$ as a function of
varying photon polarization direction from in to out of plane.
The fact that the intensities
of both these features scaled together argues against these
arising from two different bands.
Second, the very different dispersions of the peak and
hump noted in this paper also implies the same: 
if the sharp peak was a second band it must also show up 
above $T_c$, but it does not. 

We begin our discussion of self-energy effects
by recalling earlier work on the SC state lineshape.
At low energies, there is a linewidth collapse\cite{TEMP} 
upon cooling through $T_c$, leading to 
resolution limited spectral peaks\cite{NK} at $T \ll T_c$.
This is most easily understood (at least near $k_F$) 
by the freezing out of electron scattering once the SC gap
opens up.  That is, the electron lines in the Feynman diagram shown in
Fig.~3a are gapped. At higher energies ($3\Delta$ for the s-wave case)
the imaginary part of $\Sigma$ ($Im\Sigma$), 
which determines the linewidth, must go back to its normal 
state value, thus leading to a qualitative explanation \cite{THEORY}
of the dip and hump in SC state spectra near $k_F$.
For a gap with nodes, relevant to the cuprates, these arguments have 
to be modified, since contributions to $Im\Sigma$ exist
all the way to zero frequency\cite{QUIN}.  

More recently, theoretical calculations\cite{YTHEORY}, 
motivated by the neutron scattering experiments in YBCO\cite{YBCO},
have found a sharp structure in the SC state
particle-hole susceptibility (bubble in Fig.~3a).
Basically, the dominance of $Q=(\pi,\pi)$ scattering, coupled with the 
effect of the sign change of the d-wave gap with $Q$ on the BCS
coherence factors of the bubble, 
leads to a sharp resonance at $\omega_0$ between $\Delta_{max}$
and $2\Delta_{max}$ ($\Delta_{max}=\Delta_{(\pi,0)}$).
In some models there is a true collective 
mode at frequency $\omega_0$. 

Motivated by the data in Fig.~2 and the above considerations, we
incorporate the effect of a bosonic mode on the SC state spectral
function, taking proper account of both the real and the imaginary
parts of the self energy. We note that the importance of
structure in $Re\Sigma$ has not been fully appreciated in the 
context of high $T_c$ superconductors. 
A mode at $\omega_0$  leads to a
a singularity
in the SC state $Im\Sigma$ at $\omega_0+\Delta$.  By Kramers-Kronig
transformation, this implies a singularity in $Re\Sigma$
at $\omega_0+\Delta$. This feature in $Re\Sigma$ 
in turn can lead to an additional pole in
the spectral function. A simple way to visualize this is to
to think of an electron interacting
with an Einstein mode at $\omega_0$\cite{SCHRIEFFER}.  
One obtains for this model the well-known dispersion 
with two branches resulting from the mixing of the electron and the mode, 
with in general one pole being primarily 
electron-like in nature and the other being primarily mode-like. 
In addition, there is no damping for frequencies lower 
than $\omega_0+\Delta$.

We model the SC state data using the phenomenological 
self energy\cite{FOOT2}
\begin{eqnarray}
Im\Sigma(\omega)&=\Gamma_0N(|\omega|) + \Gamma_1N(|\omega|-\omega_0),
& |\omega| > \omega_0+\Delta \nonumber \\
                &=\Gamma_0N(|\omega|),
& \Delta < |\omega| < \omega_0+\Delta \nonumber \\
                &=0,  & |\omega| < \Delta
\end{eqnarray}
where $N(\omega)= \omega/\sqrt{\omega^2-\Delta^2}$ 
is the BCS density of states, and
\begin{eqnarray}
\pi Re\Sigma(\omega) = \Gamma_0N(-\omega) \ln\left[{|-\omega+
\sqrt{\omega^2-\Delta^2}|}/{\Delta}\right] \nonumber \\
 + \Gamma_1N(\omega_0-\omega) \ln\left[{|\omega_0-\omega
+\sqrt{(\omega-\omega_0)^2-\Delta^2}|}/{\Delta}\right] \nonumber \\
- \{\omega \rightarrow -\omega\}
\end{eqnarray}
The spectral function is then\cite{SCHRIEFFER}
\begin{equation}
\pi A(\omega) = Im \frac{Z\omega + \epsilon}{Z^2(\omega^2-\Delta^2)-\epsilon^2}
\end{equation}
with (a complex) $Z(\omega) = 1 - \Sigma(\omega)/\omega$.  

Note that at large $\omega$ the model has a constant linewidth broadening,
so that we are effectively ignoring the $\omega$-dependence of the normal 
state $Im\Sigma$.  
We have found that simulations with a large constant $Im\Sigma$,
which also take into account background emission, can adequately fit the
normal state data (see also \cite{OLSON}).
From normal state data, we find that $\Gamma_1$
is of order 200 meV along $\Gamma-\bar{M}$.  
The constant linewidth simplification allows us to directly obtain the
dispersion $\epsilon_{\bf k}$ from tight binding fits to  
the normal state peak position\cite{DING96}.  
Our final results are not very sensitive to $\Gamma_0$ which is included
for numerical stability; we use 30 meV, but
similar results are found for 5 meV.  (To obtain significant
damping of the low frequency pole would require making $\Delta$ complex.)
For $\Delta$, we assume a d-wave gap 
$\Delta_{\bf k} = \Delta_{max}(\cos(k_xa)-\cos(k_ya))/2$ with
$\Delta_{max}$ = 32 meV.
The best agreement with experiment is found by choosing
the bosonic mode frequency
$\omega_0 = 1.3\Delta_{\bf k}$.

The resulting real and imaginary parts of $\Sigma$ at $\bar{M}$ are shown
in Fig.~3b emphasizing the singular behaviors 
at $\Delta$ due to the $\Gamma_0$
term and at $\omega_0+\Delta$ due to the $\Gamma_1$ term.
In Fig.~4, we show a comparison of the resulting spectral
function (convolved with the experimental energy and momentum resolution)
to experimental data at $\bar{M}$ for both wide and narrow energy
scans.  To aid in the comparison to experiment, a step edge background
(which gives a good model of data for $k > k_F$) is added to the calculation.
The positions of the sharp peak and the hump obtained from
the calculations are compared to the experimental data in Fig.~2.  
The relative lack of movement of the low frequency peak is well
reproduced, as well as its lack of visibility for curves 1 through 3 
in Fig.~1b.  This lack of movement is due to several factors: 
(1) the flatness of the band dispersion $\epsilon_{\bf k}$
near $\bar{M}$, (2) the lowering of $\omega_0 \sim \Delta_{\bf k}$ 
as one moves towards $\Gamma$, and (3) the influence of both
$Re\Sigma$ and the gap.  The last is a new effect worth commenting on.
The self-energy (Eq.~2) implies a mass enhancement ($Z > 1$)
in the SC state relative to the normal state, which acts to
push spectral weight towards $E_F$.  On the other hand, $\Delta$
itself pushes spectral weight away from $E_F$.  Thus the
dispersion is dramatically flattened relative to the normal state.

We now discuss the $p$ dependence of $\Sigma$, which was essentially
ignored, except in so far as it was crudely captured by that of  
$\omega_0$. Assuming that $Q=(\pi,\pi)$ scattering dominates,
as suggested by the neutron data,  we would replace 
$\Gamma_1N(\omega+\omega_0)$ in Eq.~1 by 
$g_{p,p+Q}^2 A_{p+Q}(\omega+\omega_0)$ (for $\omega<0$)
where $g$ is the interaction vertex.
Assuming a quasiparticle pole approximation for $A$ when solving
Eqs.~1 and 2, this would imply a dip in the spectrum
at $|\omega|=E_{p+Q}+\omega_0$ where $E_p^2=\epsilon_p^2/(Re Z_p)^2
+\Delta_p^2$, and, as before, a persistent low frequency peak if $Z$ is
large enough.  Note that $p$ and $p+Q$ are coupled in the equation for
$\Sigma$ which requires a numerical solution.  Although we leave this
complexity for future work, we note that this coupling implies that if a
low frequency peak exists for $p$, then one also exists for $p+Q$ (in fact,
they self-consistently generate one another if $g$ is large enough).  
This is just the effect observed in the data along $(\pi,0)$-$(\pi,\pi)$, 
shown in Fig.~1c, in that a persistent
low frequency peak exists for about the same
angular range as that along $(\pi,0)$-$(0,0)$.  Thus the expected $Q$
dependence of the scattering is indeed reflected in the ARPES data.
We note that such scattering can also 
cause a pairing instability in the d-wave channel.

In conclusion, we have shown the presence of a persistent low frequency peak 
in photoemission spectra in Bi2212 in the SC state which exists over
a large angular range near the $\bar{M}$ point.  The dispersion of this feature
and the higher binding energy hump as a function of
momentum indicates that the electrons in the SC state are
interacting with a bosonic mode of resonant character with a frequency
between $\Delta_{max}$ and 2$\Delta_{max}$. 
Our results once again emphasize that the self-energy is dominated
by electron-electron interactions, which is consistent with an electron-electron
origin to the pairing.

We thank Yuri Vilk for several stimulating discussions.
This work was supported by the U. S. Dept. of Energy,
Basic Energy Sciences, under contract W-31-109-ENG-38, the National 
Science Foundation DMR 9624048, and
DMR 91-20000 through the Science and Technology Center for
Superconductivity.
The Synchrotron Radiation Center is supported by NSF grant DMR-9212658.

\begin{figure}
\caption{
EDCs in (a) the normal state (105 K) and (b) the superconducting state (13 K)
along
the line $\Gamma-\bar{M}-Z$, and (c) the superconducting state (13 K) along
the line $\bar{M}-Y$, for a slightly overdoped ($T_c$ = 87K)
Bi2212 sample with photon
polarization $\Gamma-\bar{M}$.  The zone is shown as an inset in (c) with
the curved line representing the observed Fermi surface.}
\label{fig1}
\end{figure}

\begin{figure}
\caption{
Positions (eV) of the sharp peak and the broad hump in the 
SC state versus normal state peak position.  
Solid points connected by a dashed line are the experimental data, 
the solid lines are obtained from the calculations described in the text,
and the dotted line represents the normal state dispersion.}
\label{fig2}
\end{figure}

\begin{figure}
\caption{
(a) Feynman diagram for the lowest order contribution to the self-energy from
electron-electron scattering.  The solids lines are electrons, the dashed lines
the interaction. (b) $Im\Sigma$ and $ReZ$ at $\bar{M}$ from Eqs.~1 and 2
for the parameter set given in the text.}
\label{fig3}
\end{figure}

\begin{figure}
\caption{
Comparison of the data at $\bar{M}$ for (a) wide and (b) narrow energy scans
with calculations based on Eqs.~1-3, with an added 
step edge background contribution.}
\label{fig4}
\end{figure}


\begin{references}
\bibitem{FOOT2}
This form is suggested by the solution of the strong coupling equation
for $\Sigma$ assuming a real, frequency independent gap.
We use an s-wave density of states to obtain an analytic result.  Results
using a d-wave density of states will be qualitatively similar.
\end{references}
\end{document}